# PERFORMANCE EVALUATION AODV, DYMO, OLSR AND ZRP Ad hoc ROUTING PROTOCOL FOR IEEE 802.11 MAC AND 802.11 DCF IN VANET USING QUALNET

Manish Sharma[1] and Gurpadam Singh[2]

[1]Department of Physics, Govt. College, Dhaliara, H.P., India
[2]Deparment of ECE, B.C.E.T.,Gurdaspur, Punjab, India

manikambli@rediffmail.com

## ABSTRACT

*In VANET high speed is the real characteristics which leads to frequent breakdown, interference etc. Therefore Performance of adhoc routing protocols is helpful to improve the Quality of Service (QOS). In this paper we studied various adhoc routing protocols, Reactive, Proactive & Hybrid, taking in to consideration parameters like speed, altitude, mobility etc in real VANET scenario. The AODV and DYMO (Reactive), OLSR (Proactive) and ZRP (hybrid) protocols are compared for IEEE 802.11(MAC) and IEEE 802.11(DCF) standard using Qualnet as a Simulation tool. Since IEEE 802.11, covers both physical and data link layer. Hence performance of the protocols in these layers helps to make a right selection of Protocol for high speed mobility. Varying parameters of VANET shows that in the real traffic scenarios proactive protocol performs more efficiently for IEEE 802.11 (MAC) and IEEE 802.11(DCF).*

## 1. INTRODUCTION

Vehicular Ad hoc Network (VANET) is a new communication paradigm that enables the communication between vehicles moving at high speeds. It has been found that mobile terminals in fast moving vehicles like cars, buses, trains are frequent signal breakdowns as compared to the lots of pedestrians. Hence in order to improve QoS and energy conservation in fast moving vehicles various light weight routing protocols needed to be studied in Physical and data link layer. So that Right selection of the protocol can be made. There are mainly three types of routing protocols, Reactive [1], Proactive [2], Hybrid [3]. These protocols have different criteria for designing and classifying routing protocols for wireless ad hoc network. The protocols in focus now days are Hybrid protocols and others [7]. Its use in the context of VANET's along with reactive and proactive is always an area under investigation. Routing protocols are always challenging in the fast moving nodes as their performance degrades with speed. Such type of networks are difficult to manage as fast handoff deteriorates signal quality, maximizes Interference and other attenuation factors. The Mobile Ad hoc Network (MANET) working group of the Internet Engineering Task Force (IETF) [4] develop standards for routing in dynamic networks of both mobile and static nodes.

Here, the feasibility, the performance, and the limits of ad hoc communication using the three types of protocols, is evaluated as per IEEE 802.11 (MAC and DCF) [5] and Potentials for optimizing the deployed transport and routing Protocols is investigated. In this work, A VANET model is designed which is based on road traffic information and includes high speed mobility at higher altitude is then fed into an event-driven network simulation. Special care is taken to provide realistic scenarios of both road traffic and network usage. This is accomplished by simulating the scenario with the help of simulation tool Qualnet [6]. The protocols and their various parameters of the transport, network, data link, and physical layers are provided by well-tested implementations of this network simulation tool.

## 2. AD HOC ROUTING PROTOCOLS

Routing protocol is a standard that controls how nodes decide how to route the incoming packets between devices in a wireless domain & further Distinguished in many types. There are mainly three types of routing protocols. Ad-hoc on demand vector distance vector (AODV), Dynamic MANET On demand (DYMO) and Dynamic source routing (DSR) are the examples of reactive routing protocols whereas Optimized Link State Routing (OLSR) and Fisheye state routing (FSR) are the examples of proactive routing protocols. Hybrid routing protocols is the combination of both proactive and reactive routing protocols, Temporary Ordered Routing Algorithm (TORA), Zone Routing Protocol (ZRP), Hazy Sighted Link State (HSLS) and Orderone Routing Protocol (OOPR) are its examples. In our work the chosen protocols are AODV, DYMO, OLSR and ZRP.

### 2.1 Ad-hoc on demand vector distance vector (AODV) Routing Protocol

AODV [8] is a reactive protocol. The reactive routing protocols do not periodically update the routing table. Instead, when there is some data to send, they initiate route discovery process through flooding which is their main routing overhead. Reactive routing protocols also suffer from the initial latency incurred in the route discovery process, which potentially makes them unsuitable for safety applications. AODV is a well known distance vector routing protocol [13] and actually works as follows. Whenever a node wants to start communication with another node, it looks for an available path to the destination node, in its local routing table. If there is no path available, then it broadcasts a route request (RREQ) message to its neighbourhood. Any node that receives this message looks for a path leading to the destination node. If there is no path then, it re-broadcasts the RREQ message and sets up a path leading to RREQ originating node. This helps in establishing the end to end path when the same node receives route reply (RREP) message. Every node follows this process until this RREQ message reaches to a node which has a valid path to the destination node or RREQ message reaches to the destination node itself. Either way the RREQ receiving node will send a RREP to the sender of RREQ message. In this way, the RREP message arrives at the source node, which originally issued RREQ message. At the end of this request-reply process a path between source and destination node is created and is available for further communication. In scenarios where there is no route error (RERR) message is issued for nodes that potentially received its RREP message. This message helps to update or recalculate the path when an intermediate node leaves a network or loses its next hop neighbour. Every node using AODV maintains a routing table, which contains the following information: a next hop node, a sequence number and a hop count. All packets destined to the destination node are sent to the next hop node. The sequence number acts as a form of time-stamping, and is a measure of the freshness of a route. This helps in using the latest available path for the communication. The hop count represents the current distance between the source and the destination node. It is important to understand that AODV does not introduce routing overhead, until a RREQ is made. This is helpful as bandwidth is not wasted unnecessarily by the routing protocol. But on the other hand this introduces an initial latency, where a node has to wait for some time to find the path to the destination and then start communication. This can be problematic for time critical and safety related emergency applications.

## 2.2 Dynamic MANET On demand (DYMO) Routing Protocol

DYMO [9] is another reactive routing protocol that works in multi hop wireless networks. It is currently being developed in the scope of IETF's [4] MANET working group and is expected to reach RFC status in the near future. DYMO is considered as a successor to the AODV routing protocols. DYMO has a simple design and is easy to implement. The basic operations of DYMO protocol are route discovery and route maintenance (I. Chakeres, & C. Perkins, 2006). When a node wants to discover a path to a destination node, it initiates the route discovery operation. A RREQ message is broadcast to the network. Every intermediate node participates in hop-by-hop dissemination of this message and records a route to the originator. When a destination node receives this RREQ message, it responds with a RREP message unicast toward the originating node. Every node receiving this message creates a route to the destination node and eventually this RREP message arrives at the originator of the RREQ message. It appears that DYMO work much like the AODV routing protocol, but there is a subtle and important difference between the two routing protocols. In addition to the route about the requested node, the originator of the RREQ message using DYMO protocol will also get information about all intermediate nodes in the newly discovered path. In AODV, only information about destination node and the next hop is maintained, while in DYMO, path to every other intermediate node is also known.

## 2.3 Optimized Link State Routing (OLSR) Protocol

OLSR [10] is the proactive routing protocol that is evaluated here. Proactive routing protocols continuously update the routing table, thus generating sustained routing overhead. Basically OLSR is an optimization of the classical link state algorithm adapted for the use in wireless ad hoc networks. In OLSR, three levels of optimization are achieved. First, few nodes are selected as Multipoint Relays (MPRs) [16] to broadcast the messages during the flooding process. This is in contrast to what is done in classical flooding mechanism, where every node broadcasts the messages and generates too much overhead traffic. OLSR achieved RFC status in 2003. Second level of optimization is achieved by using only MPRs to generate link state information. This results in minimizing the "number" of control messages flooded in the network. As a final level of optimization, an MPR can chose to report only links between itself and those nodes which have selected it as their MPR. This results in the distribution of partial link state information in the network. OLSR periodically exchanges topology information with other nodes at regular intervals. MPRs play a major role in the functionality of the protocol. Every node selects a subset of its one hop neighbour nodes as MPR. MPRs periodically announce in the network that it has reach ability to the nodes which have selected it as an MPR. Nodes which are not selected as MPR by any node, will not broadcast information received from it. The functionality of OLSR lies in the exchange of HELLO and TC messages. The periodic dissemination of HELLO packets also enables a node to know whether a node or a set of nodes have selected it as MPR. This information is known as 'Multipoint Relay Selector Set', and is critical to determine whether to broadcast forward the information received from a node(s) or not. In a dynamic and rapidly changing environment, this set of nodes can change over the time. HELLO messages are also used for link sensing and neighbourhood detection. TC messages are used to provide every node enough link-state information for the calculation of routes. Basically, a TC message is sent by a node to advertise a set of links, which includes the links to all nodes of its MPR selector set. TC message is only broadcast forwarded by MRPs and offers controlled flooding of the topology information into the whole network. OLSR is designed to support large and dense wireless networks. The levels of optimization discussed above, make it better suited for such networks.

OLSR is tailored for networks where the traffic is random and sporadic between large numbers of nodes. It is also suitable for scenarios, where the communicating pairs change over time. Once the communicating pair changes, a route to new pair is readily available, and no control traffic or route discovery process is needed as in the case of reactive protocols. This can be beneficial for situations where time critical or safety related data needs to be delivered with minimum possible delay. Due to its proactive nature, OLSR periodically generates overhead traffic. Although it is helpful in avoiding initial latency involved with route discovery, it uses precious network bandwidth for its control traffic. But it is a sustained overhead, and does not start suddenly as is the case with reactive protocols, when they start flooding the network with their control information with some application data packets waiting. Over the years, both reactive and proactive routing protocols have been used to enable communication in wireless ad hoc networks. Each approach has its own pros and cons and is suitable for its respective scenarios.

## 2.4 Zone Routing Protocol (ZRP)

Hybrid routing combines characteristics of both reactive and proactive routing protocols to make routing more scalable and efficient [11]. By and large hybrid routing protocols are zone based; it means the number of nodes is divided into different zones to make route discovery and maintenance more reliable for MANET. The need of these protocols arises with the deficiencies of proactive and reactive routing and there is demand of such protocol that can resolve on demand route discovery with a limited number of route searches. ZRP limits the range of proactive routing methods to neighbouring nodes locally; however ZRP uses reactive routing to search the desired nodes by querying the selective network nodes globally instead of sending the query to all the nodes in network. ZRP uses "Intrazone" and "Interzone" routing to provide flexible route discovery and route maintenance in the multiple ad hoc environments. Interzone routing performs route discovery through reactive routing protocol globally while intrazone routing based on proactive routing so as to maintain up-to-date route information locally within its own routing range. The overall characteristic of ZRP is reduction in the network overhead that is caused by proactive routing. It also handles the network delay that is caused by reactive routing protocols performing route discovery more efficiently. Normal routing protocols which works well in fixed networks does not show same performance in mobile ad hoc networks. In these networks routing protocols should be more dynamic so that they quickly respond to topological changes. There is a lot of work done on evaluating performance of various MANET routing protocols for constant bit rate traffic.

## 3. IEEE 802.11

IEEE 802.11[12] provides a cost effective and simple way for wireless networking. Actually IEEE has defined the specification for LAN, called IEEE 802.11, which covers both physical and data link layer [4]. The IEEE 802.11 Standard is by far the most widely deployed wireless LAN protocol. This standard species the physical, MAC and link layer operation we utilize in our simulation. Multiple physical layer encoding schemes are denned, each with a different data rate. Part of each transmission uses the lowest most reliable data rate, which is 1 Mbps. At the MAC layer[14] IEEE 802.11 uses both carrier sensing and virtual carrier sensing prior to sending data to avoid collisions. Virtual carrier sensing is accomplished through the use of Request-To-Send (RTS) and Clear-To-Send (CTS) control packets. When a node has a unicast data packet to send to its neighbour, it broadcasts a short RTS control packet. If the neighbour receives this RTS packet, then it responds with a CTS packet. If the source node receives the CTS, it transmits the data packet. Other neighbours of the source and destination that receive the RTS or CTS packets defer packet transmissions to avoid collisions by updating their network allocation vector (NAV). The NAV is used to perform virtual channel sensing by indicating that the channel is busy. After a destination properly receives a data packet, it sends an acknowledgment (ACK) to the source. This signifies other that the packet was correctly received. This procedure (RTS-CTS-Data-ACK) is called the Distributed Coordination Function (DCF). For small data packets the RTS and CTS packets may not be used. If an ACK (or CTS) is not received by the source within a short time limit after it sends a data packet (or RTS), the source will attempt to retransmit the packet up to seven times. If no ACK (or CTS) is received after multiple retries, an error is issued by the hardware indicating that a failure to send has occurred. Broadcast data packets are handled differently than unicast data packets. Broadcast packets are sent without the RTS, CTS or ACK control packets. These control messages are not needed because the data is simultaneously transmitted to all neighbouring nodes.

IEEE 802.11 also supports power saving and security. Power saving allows packets to be buffered even if the system is asleep.

### 3.1 802.11 Medium Access Control

The 802.11 MAC [12] mainly covers three functional areas Reliable Data Control, Access Control and Security. For Reliable data control it includes a frame exchange protocol. In this if source does not receive ACK (acknowledgement) within a short period of time due to data frame damaged or returning ACK damaged, the source retransmit it.

In Access Control Function there are two modes: Distributed Access protocol which like Ethernet distributes the decision to transmit over all the nodes using a carrier sense mechanism and centralized access protocol, which involve regulation of transmission by a centralized decision maker. A distributed access protocol makes sense for an adhoc wireless network .A centralized access protocol is nature for configurations in which a number of wireless stations are interconnected with one another and some sort of base stations that attaches to a backbone wired LAN, it is highly useful for data of time sensitive or high priority.

MAC is required to provide fair access to resources & efficient utilization of Bandwidth. MAC layer is also responsible for providing system authentication, association with an access point, encryption and data delivery [14].In MAC the optional priority-based point coordination function provides contention-free frame transfer for processing time-critical information transfers. With this operating mode, a point coordinator resides in the access point to control the transmission of frames from stations. When a station wants to send data, it waits for short inter-frame spacing (SIFS), and then start transmission. If there is no data to send the point coordinator waits for PCF inter-frame spacing (PIFS) and poll the next station. PCF cannot provide quality of service (QOS) mainly due to delayed contention period.

### 3.2 802.11 Distributed coordination Function

The end result for 802.11 is a MAC algorithm DFWMAC (Distributed foundation wireless MAC) that provides a Distributed access control mechanism with an optional centralized control built on top of that. The lower sublayer of the MAC layer is the Distributed coordination function. (DCF)[15] Uses a contention algorithm to provide access to all traffic. Ordinary asynchronous traffic directly uses DCF. If a station has a MAC frame to transmit it listens to the medium, if the medium is idle the station may transmit, otherwise the station must wait until the current transmission is complete. The DCF does not include collision detection function because it is not practical in a wireless network. To ensure the smooth Functioning DCF includes a set of delays that amounts to a priority scheme. The basic access mechanism, called Distributed Coordination

Function (DCF) is basically a Carrier Sense Multiple Access with Collision Avoidance mechanism (CSMA/CA). In CSMA/CA When a station wants to send, it senses the medium. If the radio channel is idle for at least the duration of DCF inter-frame spacing (DIFS), the station sends. If the medium is busy it waits until it is idle for DIFS. Then it enters a contention phase choosing randomly a multiple slot-time within a contention window (CW) as a back off timer. If the medium is idle when the timer expires, the station transmits. If the medium is busy before the timer expires, the station stops the timer and starts it again after the channel is idle for DIFS [15]. After each unsuccessful transmission attempt the size of the CW size is doubled. As a consequence, the waiting time increases and the probability of a concurrent transmission decrease. As a result of this scheme the delay is increased when a high network load is present. CSMA/CA does not overcome the hidden terminal problem, since collisions can still occur at the receiver. Therefore an extension to DCF was made, introducing two new control packets and a virtual channel reservation scheme[16] . It is called Distributed Coordination Function with RTS/CTS extension. Using this scheme a station waits for the duration of DIFS and then sends a request to send message (RTS). This includes a duration field specifying the expected time needed for the transmission of data and acknowledgement. Every node overhearing the RTS stores the medium allocation in his net allocation vector (NAV). The receiver waits for SIFS and sends a clear to send (CTS), also including the duration field. Again, all nodes over hearing the CTS set their NAV. From now the channel is reserved for the sender. He waits for SIFS and transmits the data. The receiver responds with ACK after waiting SIFS.

## 4. SIMULATION TOOL

The adopted methodology for the results of this work (specifically comparative routing analyses) is based on simulations near to the real time packages before any actual implementation. This is accomplished by simulating the scenario with the help of simulation tool Qualnet [6].QualNet is a comprehensive suite of tools for modelling large wired and wireless networks. It uses simulation and emulation to predict the behaviour and performance of networks to improve their design, operation and management. QualNet enables users to design new protocol models, Optimize new and existing models, Design large wired and wireless networks using pre-configured or user-designed models, Analyze the performance of networks and perform what-if analysis to optimize them. QualNet is the preferable simulator for ease of operation. So, we found QualNet be the best choice to implement our scenarios as we do not need every feature possible.

QualNet is a commercial simulator that grew out of GloMoSim, which was developed at the University of California, Los Angeles, UCLA, and is distributed by Scalable Network Technologies [6]. The QualNet simulator is C++ based. All protocols are implemented in a series of C++ files and are called by the simulation kernel. QualNet comes with a java based graphical user interface (GUI).It must be noted that Qualnet is a discrete event simulator which provides a good balance between ease of use and extensibility and power in terms of what scenarios can be simulated. Its considered, modular design makes it easier to modify than some other popular simulation tools. Also, it does not have as much complexity as some tools, which results in a shorter learning curve. Finally, it has quite advanced wireless modules with new technologies being incorporated into the tool relatively quickly.

Moreover QualNet 802.11 model captures most of the key aspect of 802.11. It provides support for both ad hoc and infrastructure modes of operation. Unlike some other simulators, it has a good model of the different modulation and coding schemes that are used in 802.11and suitable for both 802.11a and 802.11b[17] in this regard. A reasonably accurate model of the 802.11 Distributed Co-ordination Function (DCF) is also been implemented.

Table 1.  Simulation Parameters

| Simulator | Qualnet Version 5.o.1 |
|---|---|
| Terrain Size | 1500 x 1500 |
| Simulation time | 3000s |
| No. Of Nodes | 15 |
| Mobility | Random Way Point |
| Speed of Vehicles | Min.=3m/s Max.=20m/s |
| Routing Protocols | AODV,DYMO,OLSR,ZRP |
| Medium Access protocol | 802.11 MAC, 802.11  DCF  Tx Power=150dbm |
| Data size | 512 bytes |
| Data Interval | 250ms |
| No. of sessions | 5 |
| Altitude | 1500 |
| Weather mobility | 100ms |
| Battery model | Duracell 1500-AA |

## 5. DESIGNING OF SCENARIO

The Qualnet Simulator is used, has a scalable network library and gives accurate and efficient execution [18].The scenario is designed in such a way that it undertakes the real traffic conditions. We have chosen 15 fast moving vehicles in the region of 1500X1500 $m^2$ with the random way point mobility model. There is also well defined path for some of the vehicles. It shows wireless node connectivity of few vehicles using CBR application. The area for simulation is Hilly area with altitude of 1500 meters. Weather mobility intervals is 100ms.Pathloss model is two ray with maximum propagation distance of 100m.Battery model is Duracell 1500-AA.The simulation is performed with different node mobility speed and CBR (Constant bit rate) traffic flow.  CBR traffic flows with 512 bytes are applied. Simulations is made in different speed utilization with IEEE 802.11 Medium access control (MAC) and Distributed Coordination Function (DCF) ad hoc mode and the channel frequency is 2.4 GHz and the data rate 2mbps. The network protocol here applied is Internet Protocol version four (IPv4). By this proposed topology the failure of node can be easily detected and it gives the way for the accuracy in their performance.

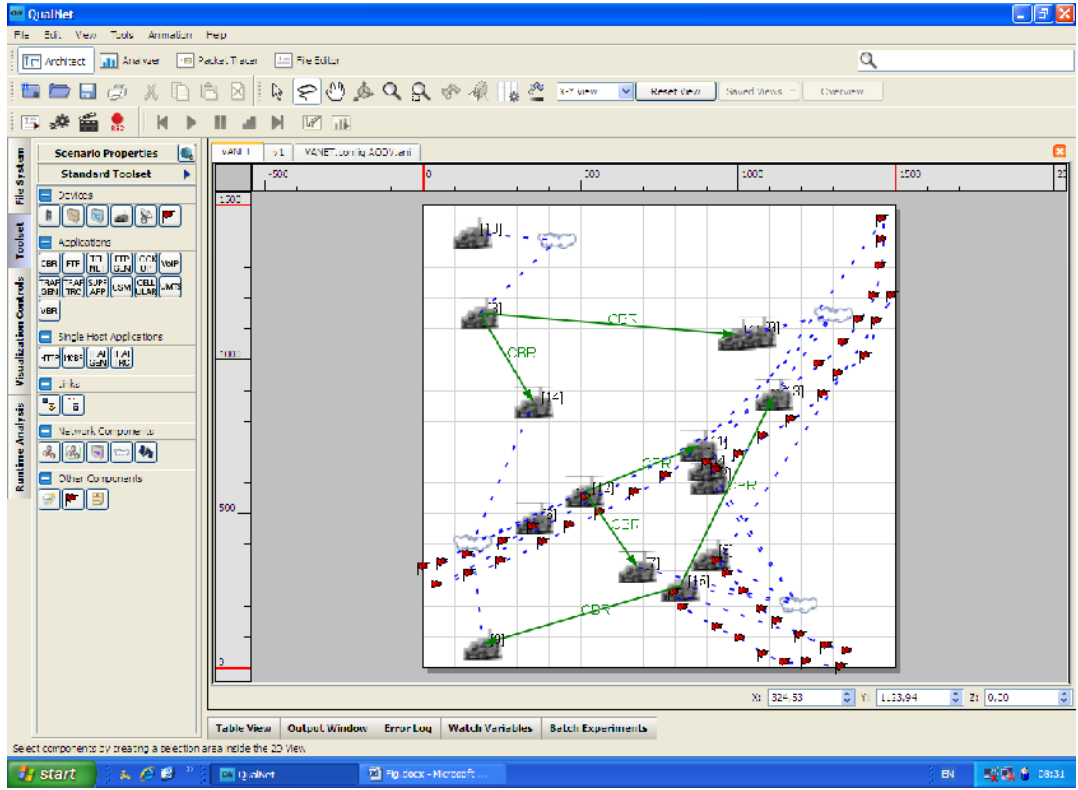

Figure 1. Qualnet VANET Scenario

## 6. RESULTS AND DISCUSSION

The simulation result brings out some important characteristic differences between the routing protocols. In all the simulation results OLSR outperforms the other protocols. This is because OLSR is a proactive protocol and it pre determines the route in well defined manner. It uses destination sequence numbers to ensure loop freedom at all times and it offers quick convergence when the network topology changes. Due to this Broadcast sent/received for IEEE 802.11 DCF is good for OLSR. In IEEE 802.11 MAC Packets from Network for OLSR protocol are very large in number as compare to others. It also shows that the Signal received without errors (802.11) and forwarded to MAC is also high. AODV and ZRP appear as the second best since both of these having some common reactive characteristics. So their performance in VANET is quite average

The least considerable performance is of DYMO protocol.Actually DYMO is different in working, although it is Reactive in nature. Besides route information about a requested target, a node will also receive information about all intermediate nodes of a newly discovered path. Therein lays a major difference Between DYMO and AODV, the latter of which only generates Route table entries for the destination node and the next hop, while DYMO stores routes for each intermediate hop. Hence its performance in IEEE 802.11 Medium access control (MAC) and Distributed Coordination Function (DCF) is least considerable, Note that residual battery capacity and Signal received with errors remains constant for all the protocols. This may be because of the small simulation time.

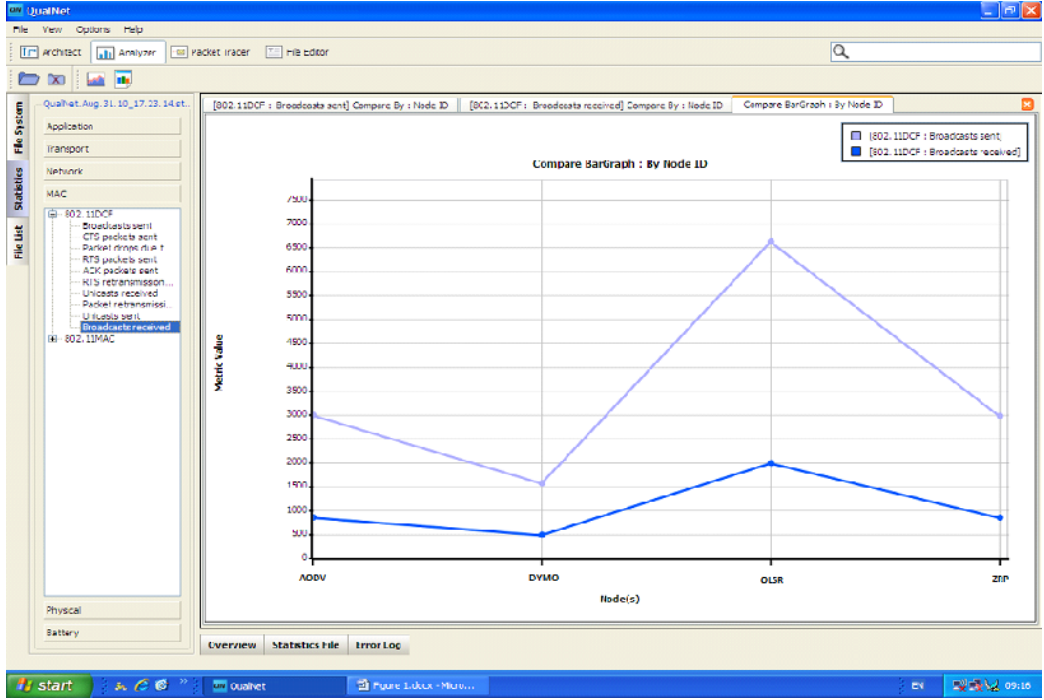

Figure 2. Broadcast sent/received (802.11 DCF) for AODV, DYMO, OLSR and ZRP

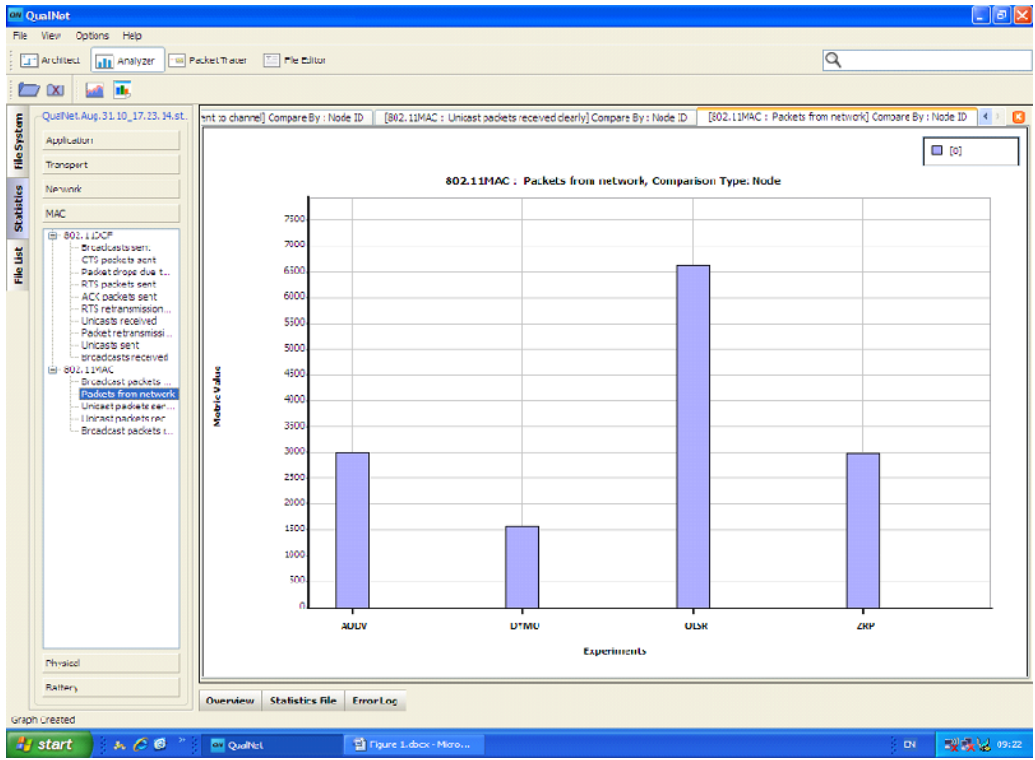

Figure 3. (802.11 MAC) Packets from Network for AODV, DYMO, OLSR and ZRP

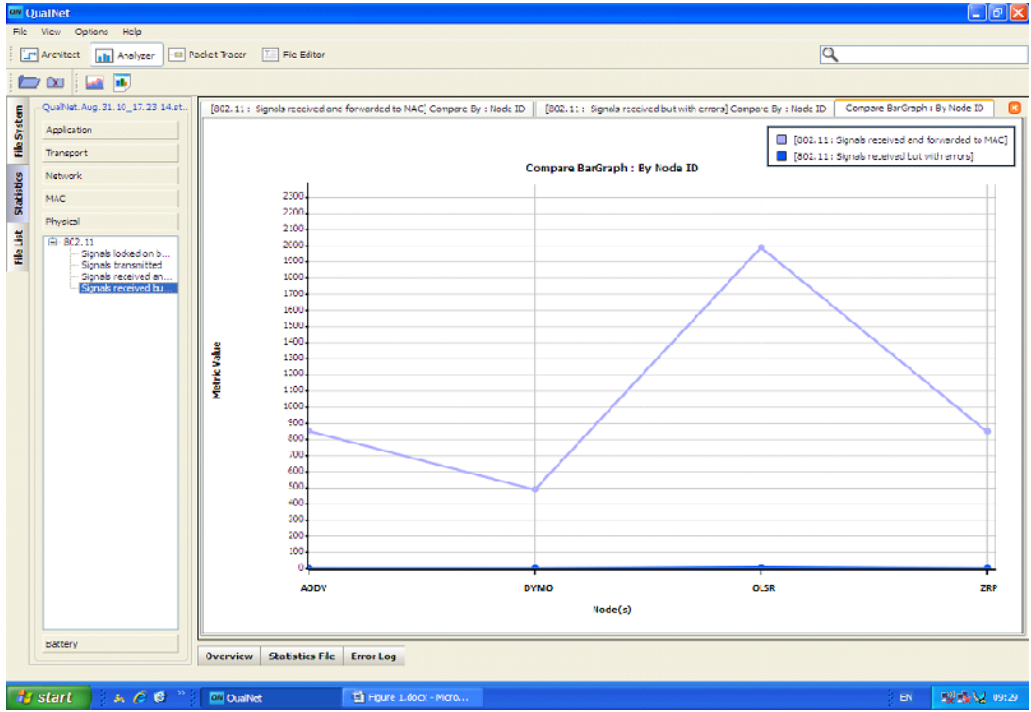

Figure 4. Signal received without/with errors (802.11) for AODV, DYMO, OLSR, and ZRP

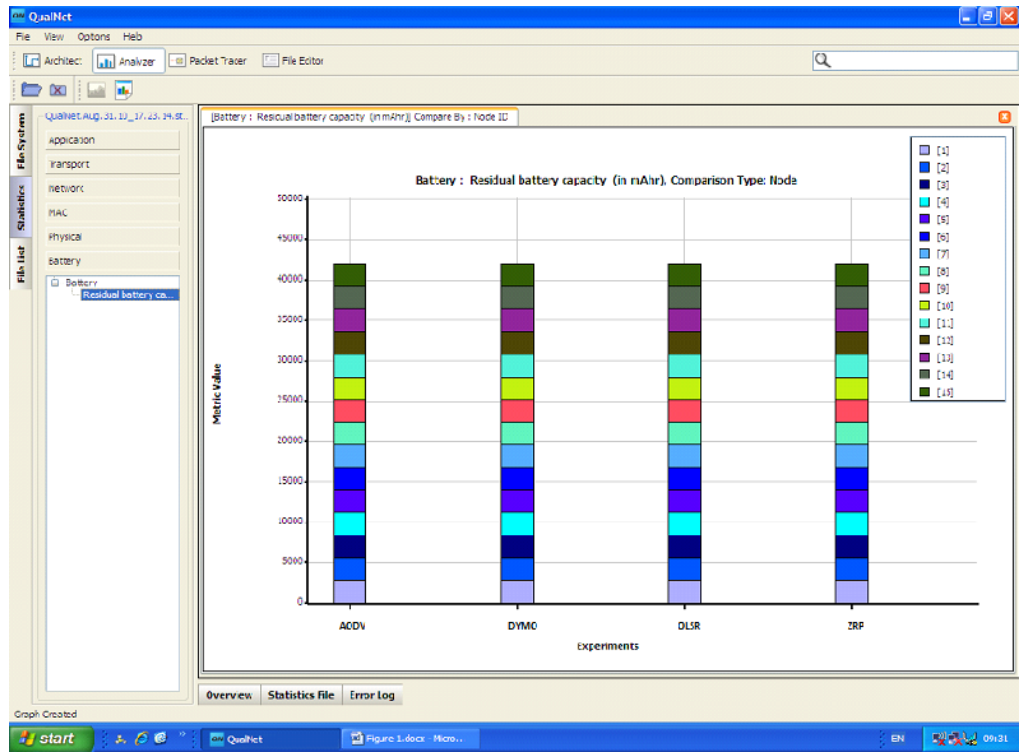

Figure 5. Residual Battery capacity for AODV, DYMO, OLSR and ZRP

## 7. CONCLUSIONS

Evaluation of the feasibility and the expected quality of VANETs operated as per IEEE 802.11 (MAC) and IEEE 802.11 (DCF) with the routing protocol OLSR at higher altitude, shows significant results. Because of its proactive nature data sent/receive is better than reactive and hybrid protocols. There are also minimum errors in signal sent. Here Adhoc networks of vehicles and static highway infrastructure can be successfully setup, maintained, and used with Qualnet simulator. Simulation results therefore seem to encourage an adaptation of the OLSR protocols for VANET use.  So that problems (frequent breakdown, interference) perceived by users are reacted to in a sensible way and application requirements are taken in to account when the network demands high QoS(quality of service). The performance of the protocols in physical, transport, data link and network layer gives important clues to improve the QoS.  In future many parameters like longitude, latitude, geographical location, traffic, can also be considered for the exact results similar to real world. Since IEEE 802.11 is heart and soul for wireless communication. Hence its performance at any level counts and shows the path for future applications.

## ACKNOWLEDGEMENTS

We are very thankful to Mr. Ashok Kumar Associate Prof. E & C Engg. NIT, Hamirpur and Dr. Nidhi Sharma, lecturer, HPES  for their support and guidance in this work.

## Authors

Manish Sharma received   M.Sc. Degree in Physics with specialisation in Electronics from H.N.B. Garhwal University in 2001 & M. Phil. in Physics with specialisation in Material Science from Annamalai University in 2006.Presently he is Pursuing his M.Tech. in Electronics & Communication from Punjab Technical University. He worked as faculty in Sri Sai College of Engg. & Tech.  Badhani   Pathankot from 2004 to 2009 & also officiated as HOD Applied Science from 2005 to 2009 in the same college.Currently he is working as Asst. Prof. in the Dept. of Physics Govt. P.G. College Dhaliara (H.P.).His Current research focuses on Wireless Communication, VANET, MANET.

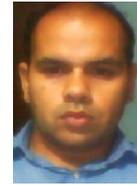

 E-mail:manikambli@rediffmail.com

Gurpadam Singh is B.E. in Electrical &    Electronics Communication & M.Tech in Electronics & Communication. Presently he is pursuing PhD in  ECE. He worked as a lecturer in GNDP Ludhiana From 1996 to 2000. And in BCET Gurdaspur from   2000 to till date. Presently he is Associate Prof & also holding key positions in the College. He is having    several National & international papers in his credits. He has also organised various seminars & Conferences & a member of ISTE.

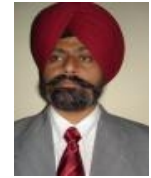

E-mail:gurpadam@yahoo.com